\documentclass[twocolumn]{aastex631}
\pdfoutput=1 
\usepackage{amsmath,amstext}
\usepackage[T1]{fontenc}
\usepackage[figure,figure*]{hypcap}
\usepackage{mwe}
\usepackage{lipsum}
\usepackage{graphicx} 
\graphicspath{{figures/}}
\usepackage{ulem}
\usepackage{multirow}
\usepackage{xcolor,colortbl}
\usepackage{tipa}
\usepackage{natbib}
\usepackage{hyperref}
\usepackage{enumitem}
\usepackage{mathtools}

\shortauthors{O'Neill et al.}
\newcommand{\alphavir}{$\alpha_{\rm{vir}}$\xspace}
\newcommand{\alphaco}{$\alpha_{\rm{vir}}(R_{CO}) $\xspace}
\newcommand{\alphaHtwo}{$\alpha_{\rm{vir}}(R_{H_2}) $\xspace}

\newcommand{\Htwo}{H$_2$\xspace }
\newcommand{\sigv}{$\sigma_v $\xspace}
\newcommand{\ts}{\textsuperscript}

\newcommand\HI{$\textrm{H}\scriptstyle\mathrm{I}$\xspace}
\newcommand{\fDG}{$f_{DG} $\xspace}

\begin{document}

\title{Effects of CO-dark Gas on Measurements of Molecular Cloud Stability \\ and the Size-Linewidth Relationship}


\author[0000-0003-4852-6485]{Theo J. O'Neill}
\affiliation{Department of Astronomy, University of Virginia, Charlottesville, VA 22904, USA}

\author[0000-0002-4663-6827]{R\'{e}my Indebetouw}
\affiliation{Department of Astronomy, University of Virginia, Charlottesville, VA 22904, USA}
\affiliation{National Radio Astronomy Observatory, 520 Edgemont Road, Charlottesville, VA 22903, USA}

\author[0000-0002-5480-5686]{Alberto D. Bolatto}
\affiliation{Department of Astronomy, University of Maryland, College Park, MD 20742, USA}

\author[0000-0003-3229-2899]{Suzanne C. Madden}
\affiliation{AIM, CEA, CNRS, Université Paris-Saclay, Université Paris Diderot, Sorbonne Paris Cité, 91191, Gif-sur-Yvette, France}

\author[0000-0002-7759-0585]{Tony Wong}
\affiliation{Department of Astronomy, University of Illinois, Urbana, IL 61801, USA}

\begin{abstract}
Stars form within molecular clouds, so characterizing the physical states of molecular clouds is key in understanding the process of star formation.  Cloud structure and stability is frequently assessed using metrics including the virial parameter and Larson (1981) scaling relationships between cloud radius, velocity dispersion, and surface density.  Departures from the typical Galactic relationships between these quantities have been observed in low metallicity environments.  The amount of H$_2$ gas in cloud envelopes without corresponding CO emission is expected to be high under these conditions; therefore, this “CO-dark” gas could plausibly be responsible for the observed variations in cloud properties.  We derive simple corrections that can be applied to empirical clump properties (mass, radius, velocity dispersion, surface density, and virial parameter) to account for CO-dark gas in clumps following power-law and Plummer mass density profiles.  We find that CO-dark gas is not likely to be the cause of departures from Larson’s relationships in low-metallicity regions, but that virial parameters may be systematically overestimated.  We demonstrate that correcting for CO-dark gas is critical for accurately comparing the dynamical state and evolution of molecular clouds across diverse environments.
\end{abstract}

\section{Introduction}

Star formation is strongly correlated with tracers of molecular gas over kpc-scales \citep[e.g.,][]{Kennicutt2007,Leroy2008,Bigiel2011}, suggesting a causal relationship between the two.  Since molecular clouds are the sites of star formation, understanding their dynamical states is necessary in accurately predicting star formation both in individual clouds as well as across larger populations.

Molecular hydrogen \Htwo is the most abundant molecule in the interstellar medium (ISM) and is therefore closely tied to understanding the stability of molecular clouds and process of star formation.  \Htwo is a symmetric, homonuclear molecule with widely spaced rotational energy levels and no permanent dipole moment; as a consequence of this, it radiates very weakly and is difficult to observe directly under conditions typical of molecular clouds ($T \sim$ 10--20 K).  It is therefore necessary to use more accessible molecules as tracers of \Htwo to fully understand the conditions under which stars form. 

CO is one of the next most abundant molecules in the ISM and can be excited easily at low temperatures, making it a popular tracer of H$_2$.  Using CO as a tracer, the amount and spatial distribution of molecular gas in a region is often used to infer the process of star formation; however, this use of CO as a proxy for \Htwo relies on the assumption that it faithfully traces the full spatial extent of H$_2$.  It is well known that some portion of the \Htwo in molecular clouds is not traced by CO: since CO is less efficient at shielding itself from FUV radiation than \Htwo is, the transition from C$^{+}$ to CO occurs closer to the center of clouds than the transition from \HI{} to H$_2$, 
resulting in a central CO-traceable region surrounded by an extended diffuse envelope of ``CO-dark'' H$_2$.

\begin{figure}
    \centering
    \includegraphics[width=0.45\textwidth]{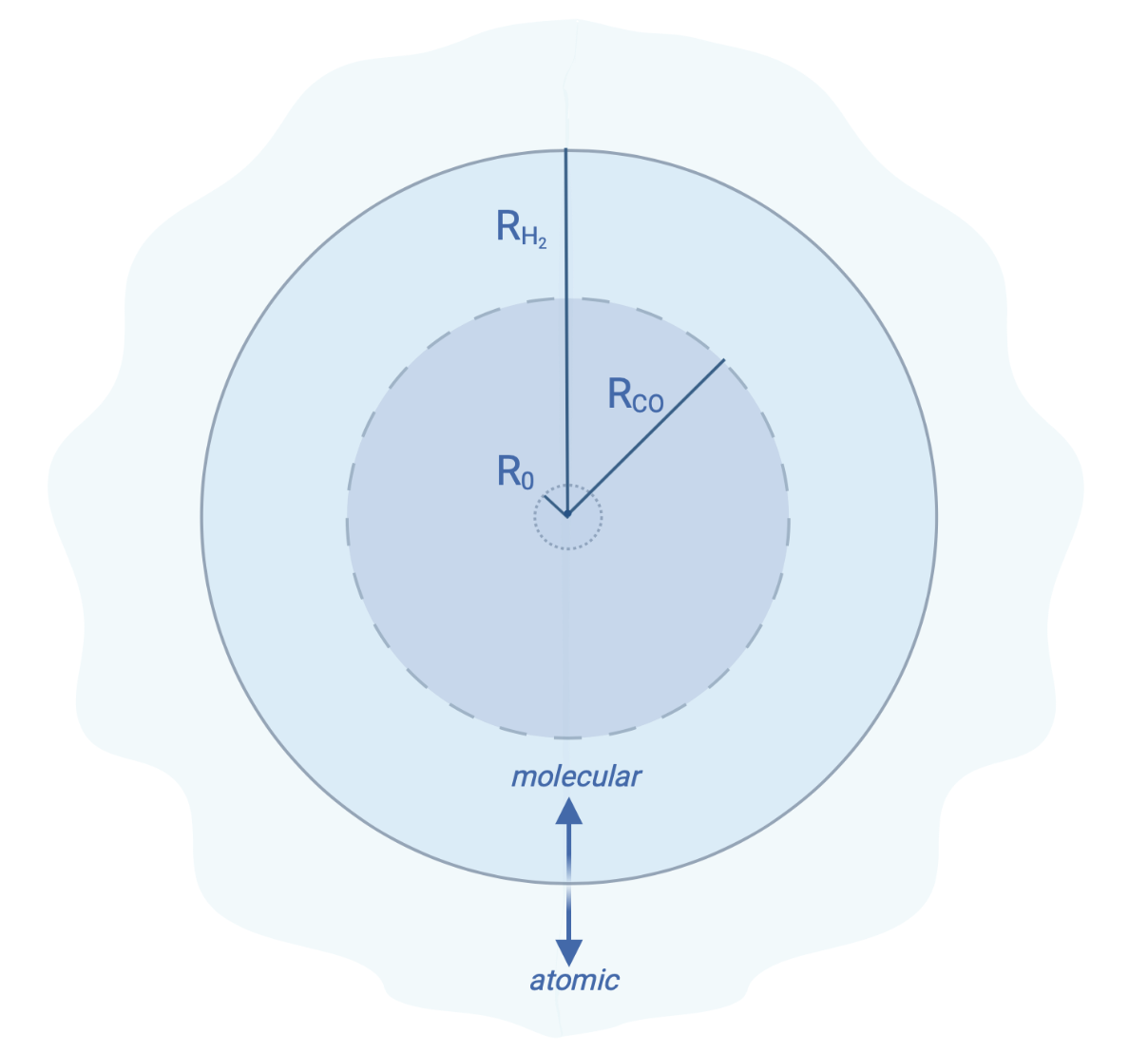}
    \caption{Clump toy model adapted from \citet{wolfire_dark_2010}.  $R_{H_2}$ is the radius at which the densities of atomic and molecular hydrogen are equal. Gas within $R_{H_2}$ is mostly molecular and gas outside of $R_{H_2}$ is mostly atomic. $R_{CO}$ is the radius at which CO-traced material has $\tau=1$ in the J=1--0 transition and is a function of $f_{DG}$, with higher $f_{DG}$ yielding smaller $R_{CO}$.  $R_{0}$ is the normalizing radius and is typically typically $\ll R_{CO}$. For a pure power-law density profile (\S\ref{S:power}) $R_0$ is arbitrary, while for a power-law profile with a core (\S\ref{S:powercore}) or Plummer profile (\S\ref{S:plummer}) it represents the radius of the flat central core.}
    \label{fig:cloudmodel}
\end{figure}

Recent studies have simulated the formation of \Htwo and CO in the ISM to evaluate the expected amount of CO-dark \Htwo in a variety of environments \citep[e.g,][]{glover_federrath_2010,glover_relationship_2011,li_2018,gong2018}.  \citet[][hereafter W10]{wolfire_dark_2010} modeled photodissociation regions (PDRs) of individual spherical clouds and defined the fraction of molecular H$_{2}$ mass not traced by CO, or ``dark gas fraction,'' as
\begin{equation}
    f_{DG} = 1 - \frac{M(R_{CO})}{M(R_{H_2})}, 
    \label{eqn:fdg_Gen}
\end{equation}
where $M(r)$ represents the mass contained within a radius $r$, $R_{CO}$ is the radius of the CO-traceable material at which the optical depth, $\tau$, equals 1 in the $J$ = 1--0 transition, 
and $R_{H_2}$ is the radius at which half of the hydrogen in the envelope surrounding the CO clump is molecular and half is atomic.  This model is shown in Figure \ref{fig:cloudmodel}. 
Assuming standard Galactic conditions, W10 derived $f_{DG} \sim$ 0.3, a result which they found to be relatively insensitive to cloud and environmental properties.  Other studies both of individual cloud envelopes and at galactic scales have derived $f_{DG} \sim 0.4$, but observed a stronger dependence on environmental properties \citep[e.g.,][]{smith2014,szucs_how_2016}.

Similar values of \fDG have been found through observational work.  In studies of individual Galactic clouds, $f_{DG}$ has been found to be $\gtrsim$0.3 \citep[e.g.,][]{grenier2005,Abdo2010,velusamy2010,lee2012,langer2014,xu2016}, and on galactic scales CO-dark gas has been inferred to be 0.2--0.3x as massive as the total atomic mass of the Milky Way and 1.2--1.6x as massive as its total CO-traced molecular mass \citep{planck2011,Paradis2012}.

The amount of CO-dark gas is expected to increase in high-radiation environments, with the C\ts{+}/C\ts{0}/CO transition shifting even further into the cloud to reach higher overall column densities.  
Similarly, the dark-gas fraction is expected to increase in low-metallicity (low-Z) environments, where decreasing dust-to-gas ratios  combine with typically stronger radiation fields to increase the efficiency of CO destruction \citep{madden_ism_2006,gordon_surveying_2011,madden2020}.  
H$_2$ can additionally be photodissociated via Lyman-Werner band photons, but since it can be optically thick under some $A_V$ conditions it is able to  remain self-shielded while CO is photodissociated.  These effects have been supported observationally in the metal-poor outskirts of the Galaxy and in the Large and Small Magellanic Clouds (LMC and SMC, respectively, with $Z\sim$1/2 $Z_\odot$ and $Z\sim$1/5 $Z_\odot$) where $f_{DG}\gtrsim $ 0.8 \citep{pineda2013,jameson_first_2018,chevance_co-dark_2020}.

Although much work has gone into quantifying the cause and amount of CO-dark gas in a variety of environments, the practical impact of this gas on interpretations of metrics of clump stability and evolution has not been explored in as much depth.  Assessing the gravitational stability of clouds as measured by the virial parameter \alphavir \citep{bertoldi_pressure-confined_1992} or if clouds conform to ``Larson's {\color{black}relationships'' between} cloud radius,  velocity dispersion, and surface density \citep{larson1981} is ubiquitous in both theoretical and observational studies.  In low-Z environments, departures from the typical values and relationships between these quantities for CO clouds under Galactic conditions have been observed \citep[e.g.,][]{bolatto2008,Hughes2013,rubele_vmc_2015,ochsendorf_what_2017,kalari_resolved_2020}.  CO-dark gas could plausibly be responsible for these variations, since \fDG is known to be high in these regions and cloud properties inferred from CO-traced material are not guaranteed to be representative of the overall state of the structures. 
Correcting for CO-dark gas may then be an essential step in evaluating the dynamical states and likely futures of molecular clouds across a range of environments.  

Here we present explicitly the variation in cloud properties from what would be inferred using CO-traceable material to the ``true'' state of clouds including CO-dark gas.  In \S\ref{S:density} we summarize mass density profiles that clouds may follow, derive corrections for empirical clump properties to account for CO-dark gas, and explore the behavior of \alphavir as $f_{DG}$ increases.  We demonstrate the biases CO-dark gas creates in interpretations of {\color{black}size-linewidth-surface density scaling relationships} in \S\ref{S:effects}.  We discuss the implications of our results and the effects of CO-dark gas on star formation in \S\ref{S:discuss} before concluding in \S\ref{S:conclude}.

\section{Clump Density Profiles}\label{S:density}

\begin{figure*}
    \centering
    \includegraphics[width=\textwidth]{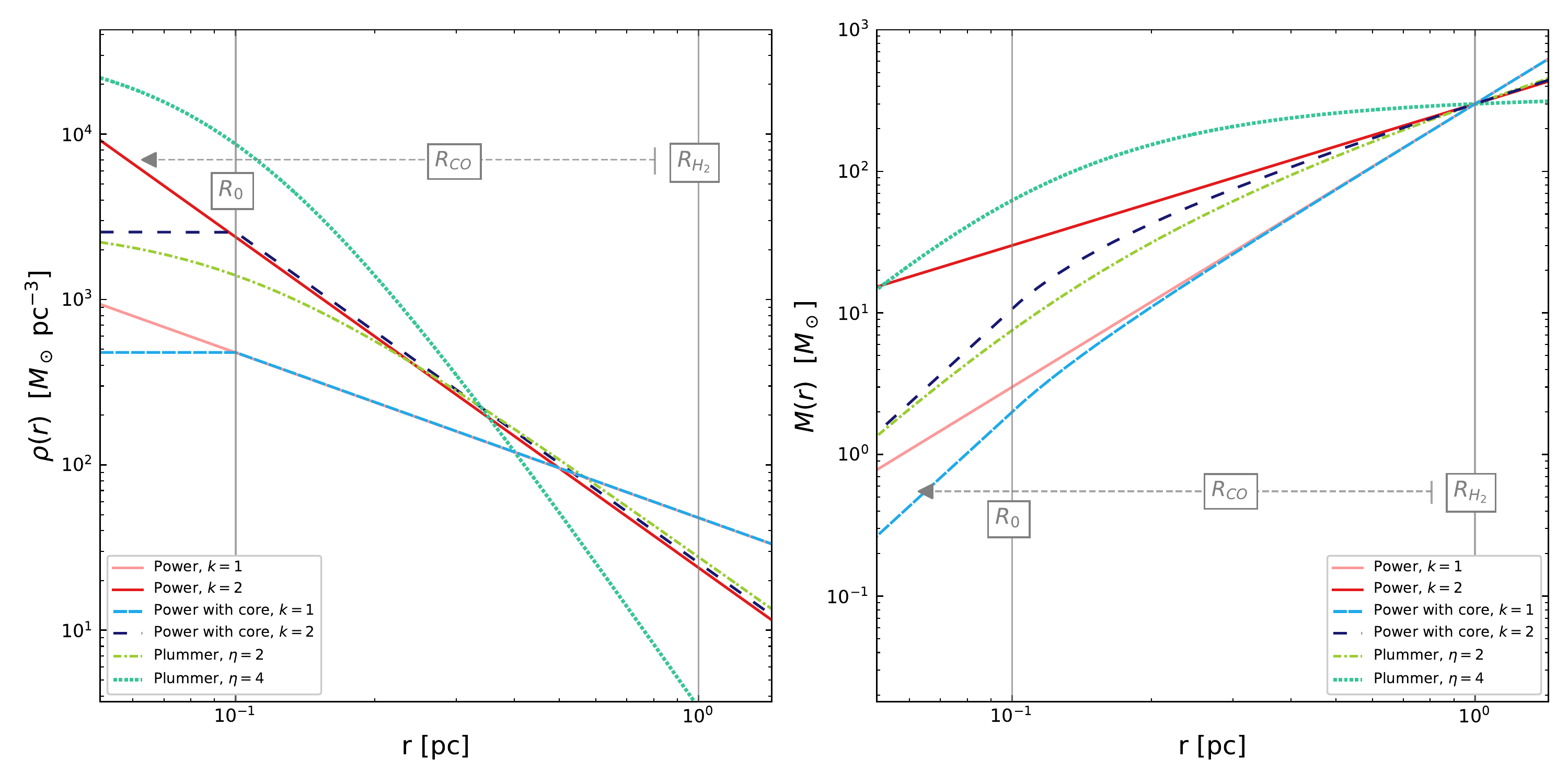}
    \caption{\textit{Left:}  Density as a function of radius for a clump with [$R_0=0.1$ pc, $R_{H_2}=1$ pc, $M(R_{H_2})=300 \ M_\odot$, $\sigma_v(R_{H_2})=$ 0.6 km  s\ts{-1}]. $R_0$ and $R_{H_2}$ are marked by the solid grey vertical lines, and the range of possible $R_{CO}$ is shown by the dashed grey arrow.  {\color{black}The solid pink and red curves are power-law profiles with $k=1$ and $k=2$, respectively (\S\ref{S:power}).  The light blue densely-dashed and dark blue loosely-dashed curves are $k=1$ and $k=2$ power-laws  with constant density cores, respectively (\S\ref{S:powercore}).  The dash-dotted yellow-green and dotted light green curves are Plummer density profiles with $\eta=2$ and $\eta=4$, respectively (\S\ref{S:plummer}).}  \textit{Right:} Total mass within $r$ as a function of $r$ for the fiducial clump, with the same line colors and styles as on the left.}
    \label{fig:profs}
\end{figure*}

\begin{figure*}
    \centering
    \includegraphics[width=\textwidth]{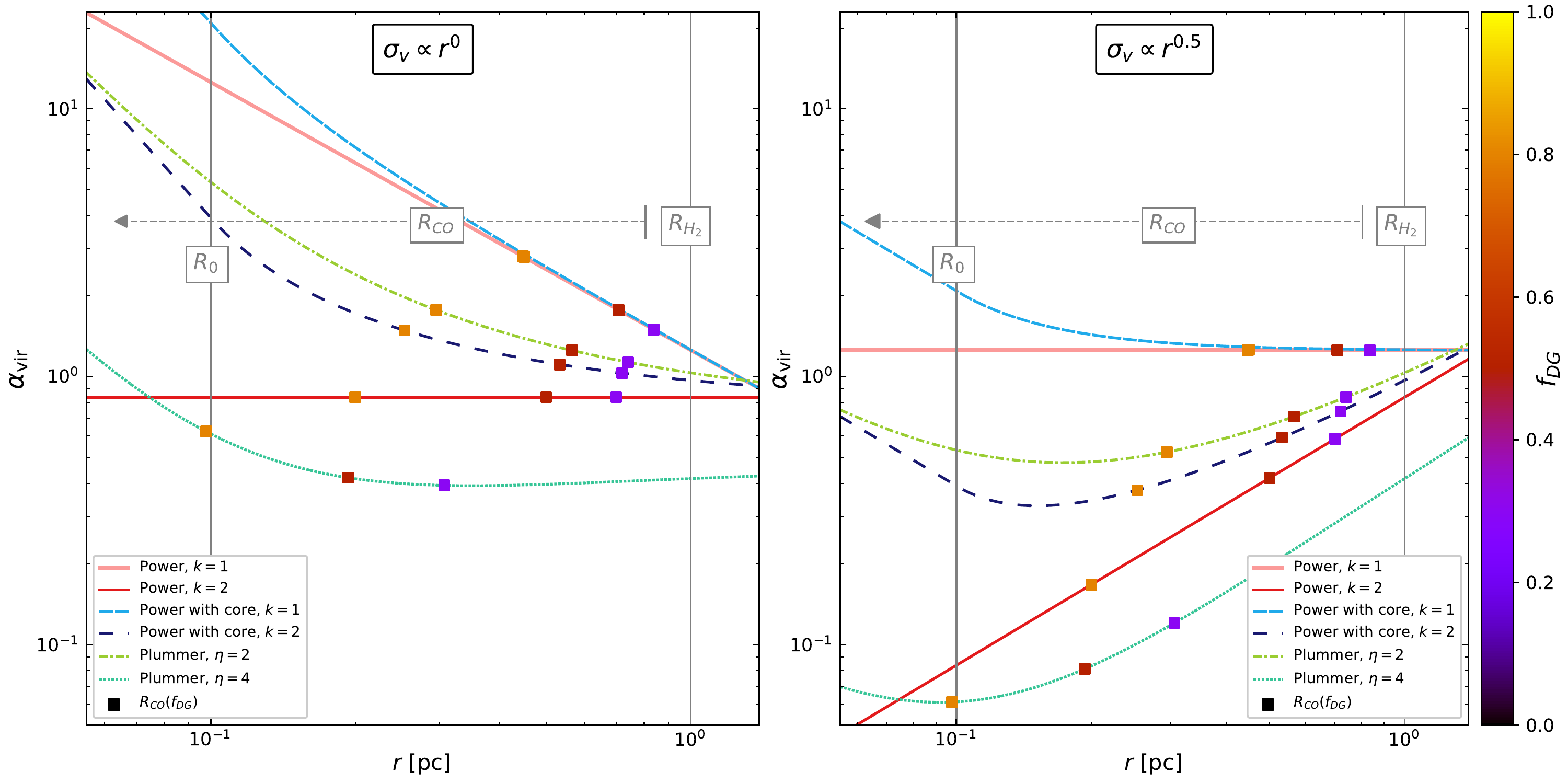}
    \caption{\textit{Left:} The value of \alphavir vs $r$ for $\sigma_v \propto r^0$ [$\beta=0$] for the clump shown in Figure \ref{fig:profs} with [$R_0=0.1$ pc, $R_{H_2}=1$ pc, $M(R_{H_2})=300 \ M_\odot$, $\sigma_v(R_{H_2})=$ 0.6 km  s\ts{-1}].  The colors and styles of the profile curves, {\color{black}the $R_0$ and $R_{H_2}$ lines, and the $R_{CO}$ arrow} are the same as in Figure \ref{fig:profs}.  The square points along each curve mark the location of $R_{CO}$ for that profile by value of $f_{DG}$.  The purple points mark $f_{DG}=0.3$, the red points mark $f_{DG}=0.5$, and the orange points mark $f_{DG}=0.8$.   \textit{Right:} Same as the left, but for $\sigma_v \propto r^{0.5}$ [$\beta=0.5$].}
    \label{fig:alphavir}
\end{figure*}

Analyzing the stability of molecular clouds ($R \gtrsim 10$ pc), clumps ($R \sim 1$ pc), and cores ($R \lesssim 0.1$ pc) is of great interest to studies of their likely evolutionary futures.  To this end, \citet{bertoldi_pressure-confined_1992} defined the virial parameter,
\begin{equation}
\alpha_{\rm{vir}} = \frac{2 \Omega_K}{| \Omega_G|} = \frac{M_{\rm{vir}}}{M},
\label{eqn:alvir}
\end{equation}
as a measure of stability, where $\Omega_K$ is the kinetic energy, $\Omega_G$ is the gravitational potential energy, $M$ is the structure's mass, and $M_{\rm{vir}}$ is its virial mass.  \alphavir $<$1 suggests that the structure is gravitationally dominated and rapidly collapsing, \alphavir $\sim 1$ indicates that a structure is gravitationally stable, and \alphavir $\gg 1$ suggests that a structure is sub-critical and will likely expand unless confined by external pressure.

Variations from the expected equilibrium values of \alphavir have been observed in environments where \fDG is known to be high.  In Galactic environments, \alphavir is frequently $\lesssim$2 \citep[see][for a review]{kauffmann_low_2013} in clumps and clouds.  In nearby low-Z dwarf galaxies and low density, low-pressure environments, \alphavir is frequently observed to be much larger and can reach measured values of 4--10 or more \citep[e.g.,][]{schruba_physical_2017,  schruba_how_2019}.  Since these environments are rich in CO-dark gas, it is possible the measured \alphavir could be unrepresentative of the states of full clumps, and that this additional molecular reservoir is responsible for the variations in measured \alphavir.  { \color{black} Alternatively, these differences could also be explained by measurement errors in \sigv and R stemming from large distance uncertainties, low velocity resolutions, or varying definitions of cloud radius.}

{\color{black}For all clump density profiles that we will consider, we assume a one-dimensional radial velocity dispersion profile $\sigma_v(r)$ of
\begin{equation}
    \sigma_{v}(r) = \left(\frac{r}{R_0}\right)^{\beta} \  \sigma_{v}(R_0),
    \label{eqn:sigmav}
\end{equation}
where $R_0$ is a normalizing radius} as shown in Figure \ref{fig:cloudmodel}.  {\color{black}When considered} in combination with a non-constant density profile $\rho(r)$, and if one considers turbulence to act as pressure support, {\color{black}our adopted Equation \ref{eqn:sigmav}} leads to a gradient in energy density $\sim \rho \sigma_v^2$; we address the implications of this effect for cloud stability in \S\ref{S:discuss}.  Additionally, we recognize that at very small scales ($\sim$0.1 pc) the effective pressure profile changes from thermal to non-thermal dominated support.  Since the bulk of this work considers the effects of CO-dark gas on parsec-scales, this behavior should not impact our conclusions.

We analyze how the observationally-derived \alphavir depends on observed CO radius for clouds following a single power-law (\S\ref{S:power}), a power-law with a constant density core (\S\ref{S:powercore}), and a Plummer profile (\S\ref{S:plummer}).  We derive corrections for empirical clump properties {\color{black}at a given dark-gas fraction $f_{DG}$.  \color{black} Finally, in \S\ref{S:comp_profs} we compare the behaviors of the profiles considered and discuss the impact of density profile on the effects of CO-dark gas.}

\subsection{Power-law Profile}\label{S:power}

Clouds are very frequently modeled as having a density profile $\rho(r)$ following a simple power-law, 
\begin{equation}
    \rho(r) = \rho_c \ x^{-k},
    \label{eqn:rho_plaw}
\end{equation}
where $\rho_c$ is the central density, $x=r/R_0$, and $R_0$ is an arbitrary radius at which $\rho$ is normalized.  Figure \ref{fig:profs} shows $\rho(r)$ and mass $M(r)$ as a function of $r$ for a clump with properties [$R_0=0.1$ pc, $R_{H_2}=1$ pc, $M(R_{H_2})=300 \ M_\odot$, and $\sigma_v(R_{H_2})=$ 0.6 km  s\ts{-1}] following $k=1$ and $k=2$.  

{\color{black} In Appendix \ref{ap:power}, we derive the virial parameter} for a clump following a power-law profile,
\begin{equation}
\begin{split}
    \alpha_{\rm{vir}}(r) & = \frac{3 \sigma_v^2(r)}{\pi \rho_c G R_0^2} \frac{T_1(r)}{T_2(r)},
\end{split}
\end{equation}
where $T_1(r) = \left[\frac{4x^{(3-k)}}{(3-k)}\right]$ and $T_2(r) = \left[\frac{16 x^{(5-2k)}}{(5-2k)(3-k)}\right]$.  Figure \ref{fig:alphavir} shows the variation of \alphavir with $r$ for the $k=1$ and $k=2$ profiles of the clump shown in Figure \ref{fig:profs}.  We observe a large range of outcomes as $r$ increases depending on the velocity and power-law indices adopted.  For [$k=1$, $\beta=0$], the cloud has decreasing \alphavir value as radius increases; while for [$k=1$, $\beta=0.5$], \alphavir is constant.  Similarly for [$k=2$, $\beta=0$], \alphavir is constant, and for [$k=2$, $\beta=0.5$], \alphavir increases with radius. 

We then cast these equations in terms of W10's \fDG for more insight {\color{black}and to derive corrections to observed molecular cloud properties for CO-dark gas.  For a cloud following a power-law profile with $k < 3$}, W10 defined
\begin{equation}
f_{DG}  = 1 - \left( \frac{R_{CO}}{R_{H_{2}}} \right)^{3-k}. 
\label{eqn:fdg_gen}
\end{equation}
We derive the variation in clump properties as a function of \fDG.  Using the definition of \fDG in Equation \ref{eqn:fdg_Gen}, the total molecular mass within $R_{H_{2}}$ can be found as
\begin{equation}
    M(R_{H_2}) = \frac{M(R_{CO})}{1-f_{DG}}.
    \label{eqn:mDG}
\end{equation}
From Equation \ref{eqn:fdg_gen} the relationship between $R_{CO}$ and $R_{H_2}$ is dependent on the adopted $k$,
\begin{equation}
    R_{H_2} =  (1-f_{DG})^{1/(k-3)} \ R_{CO}
    \label{eqn:rh2_plaw},
\end{equation}
and with Equation \ref{eqn:sigmav} evaluated at $R_0=R_{CO}$, $\sigma_v(R_{H_2})$ can be found as 
\begin{equation}
\sigma_v(R_{H_2}) = (1-f_{DG})^{\beta / (k-3)} \ \sigma_v(R_{CO}).
\label{eqn:fdg_sigv}
\end{equation}
Using Equations \ref{eqn:fdg_gen} and \ref{eqn:mDG}, surface density $\Sigma(r)= M(r) / \pi r^2$ at $R_{H_2}$ becomes
\begin{equation}
    \Sigma(R_{H_2})  =  (1-f_{DG})^{(1-k)/(k-3)} \  \Sigma(R_{CO}).
    \label{eqn:fdg_surfp}
\end{equation}
The virial mass  can be expressed in terms of \fDG as
\begin{equation}
    M_{\rm{vir}}(R_{H_2}) =   (1-f_{DG})^{(2\beta + 1)/(k-3)} \ M_{\rm{vir}}(R_{CO}).
\end{equation}
Finally, the CO-dark-corrected virial parameter is 
\begin{equation}
\begin{split}
    \alpha_{\rm{vir}}(R_{H_2}) = \alpha_{\rm{vir}}(R_{CO}) \ (1-f_{DG})^{(2\beta + k - 2)/(k-3)}.
\end{split}
\label{eqn:fdg_cor_avir}
\end{equation}

In Figure \ref{fig:alphavir} we show the values of $R_{CO}$ and $\alpha_{\rm{vir}}(R_{CO})$ as a function of \fDG for the fixed $R(H_2)$=1 pc clump: for \fDG = 0.3, $R_{CO}\simeq$ [0.85 pc for $k=1$, 0.7 pc for $k=2$], while for \fDG=0.5, $R_{CO}\simeq$ [0.7 pc for $k=1$, 0.5 pc for $k=2$], and for \fDG=0.8, $R_{CO}\simeq$ [0.45 pc for $k=1$, 0.2 pc for $k=2$].  

{\color{black}We consider internal pressure for the power-law profile.}  Under a polytropic model, turbulent pressure within a cloud is described by $P \sim \rho \sigma_v^2$.  By Equations \ref{eqn:sigmav} and \ref{eqn:rho_plaw}, the pressure gradient for this profile then follows d$P$/d$x$ $\sim (2\beta - k)x^{2\beta-k-1}$.  Thus, if $2\beta - k < 0$ an outward pressure gradient conducive to stability will be present throughout the clump.  Additionally, from Equation \ref{eqn:fdg_cor_avir} we see that while $2\beta + k < 2$, \alphaHtwo$<$ $\alpha_{\rm{vir}}(R_{CO})$, i.e., the empirical \alphavir from the CO-traced clump would overestimate the ``true'' \alphavir of the full cloud including CO-dark gas.  In this case, relying on the CO-derived measurement alone would lead to the incorrect conclusion that the cloud is dominated by kinetic energy, and either unbound or confined by high levels of external pressure.  

\subsection{Power-law Profile with a Constant Density Core}\label{S:powercore}

We also examine a cloud profile that follows a power-law at large $r$ but has a small, constant density core of radius $R_0$ at its center, 
\begin{equation}
\rho(r) = \begin{cases} \rho_c \ \rm{for} \ r < R_0 \\ 
\rho_c \ x^{-k} \ \rm{for} \ r \geq R_0,
\end{cases}
\end{equation}
where {\color{black}$\rho_c$ is the central density and} $x = r/R_0$.  This has frequently been supported observationally, with $R_0 \lesssim 0.1$ pc \citep[e.g.,][]{girichidis11,Juvela2018,Tang2018}.  
We note that for this profile, unlike for the full power-law profile of \S\ref{S:power}, $R_0$ has a definite physical meaning, and that $R_0$ is typically $\ll R_{CO}$.  {\color{black} In Figure \ref{fig:profs}, $\rho(r)$ and $M(r)$ are shown for $k=1$ and $k=2$} for a clump with an identical set of properties at $R_{H_2}$ to the clump {\color{black} considered in} \S\ref{S:power} [$R_0=0.1$ pc, $R_{H_2}=1$ pc, $M(R_{H_2})=300 \ M_\odot$, $\sigma_v(R_{H_2})=$ 0.6 km  s\ts{-1}].  The densities of this profile and of the full power-law profile described in \S\ref{S:plummer} are roughly in agreement at about 0.5 pc ; this is a consequence of the choice of $R_0$ and $R_{H_2}$, and changing their values changes this radius of agreement.   {\color{black}In Appendix \ref{ap:core}, we follow the process outlined in \S\ref{S:power} to derive the virial parameter for this profile,}
\begin{equation}
\alpha_{\rm{vir}}(r) = \begin{cases}
\frac{15 \sigma_v^2(r)}{4 \pi \rho_c G R_0^2} \frac{1}{x^2} \ \rm{for} \ r < R_0 \\
\frac{3 \sigma_v^2(r)}{\pi  \rho_c G R_0^2} \frac{\Pi_1(r)}{\Pi_2(r)} \  \rm{for} \ r \geq R_0.
\end{cases}
\end{equation}
where $\Pi_1(r) = \left[\frac{4}{3-k} \left(x^{3-k} - \frac{k}{3}\right) \right]$ and
\begin{equation}
    \Pi_2(r) = 
    \begin{cases}
    {\color{black}\frac{16}{3-k} \left(\frac{x^{5-2k}-1}{5-2k} + \frac{k(1-x^{2-k})}{6-3k} + \frac{3-k}{15} \right) \  \rm{for} \ k \neq 2} \\[0.5em]

    {\color{black}\frac{16}{3-k} \left( \frac{x^{5-2k}-1}{5-2k} - \frac{k \ln(x)}{3} + \frac{3-k}{15}    \right) \  \rm{for} \ k = 2. }
    \end{cases}
\end{equation}

The variation of \alphavir with $r$ is shown in Figure \ref{fig:alphavir} for $k=1$ and $k=2$.    
We observe a wide variety of behaviors as the area considered outside of the central core $R_0$ increases depending on the assumed density and velocity profiles.  For both [$k$=1, $\beta$=0] and [$k=2$, $\beta$=0], \alphavir decreases rapidly with increasing $x$.  For [$k=1$, $\beta=0.5$], \alphavir plateaus marginally below the value of \alphavir at $x=1$, and for [$k=2$, $\beta=0.5$], it increases rapidly.  Any conclusions as to whether the virial parameter of the CO-traceable material accurately represents the entire cloud, including CO-dark gas, are then extremely dependent on the assumptions made.

{\color{black}In Appendix \ref{ap:core}, we also derive} 
\begin{equation}
    f_{DG} = 
    \begin{cases}
    {\color{black}1 - \left(\frac{3-k}{3}
    \frac{(R_{CO}/R_0)^3}{ (R_{H_2}/R_0)^{(3-k)} - \frac{k}{3} } \right) \  \rm{for} \ R_{CO} < R_0} \\[1em] 
    1 - \left(\frac{(R_{CO}/R_0)^{(3-k)} - \frac{k}{3}}{(R_{H_2}/R_0)^{(3-k)} - \frac{k}{3}}\right) \  \rm{for} \ R_{CO} \geq R_0,
    \end{cases}
    \label{eqn:ap_core_fdg}
\end{equation}
assuming $R_{H_2} > R_0$.  {\color{black}We show the value of $R_{CO}$ as a function of \fDG in Figure \ref{fig:alphavir} for \fDG = 0.3, 0.5, and 0.8.  For \fDG = 0.3, $R_{CO}\simeq$ [0.85 pc for $k=1$,  0.7 pc for $k=2$], while for \fDG=0.5, $R_{CO}\simeq$ [0.7 pc for $k=1$, 0.5 pc for $k=2$], and for \fDG=0.8, $R_{CO}\simeq$ [0.45 pc for $k=1$, 0.25 pc for $k=2$].} 

We consider the limit of a clump with a very large central core, such that $R_0$ approaches $R_{H_2}$ and $k$ is effectively zero throughout the clump.  In this case, we can derive a simplified kinetic term, $\Omega_k \propto x^{2\beta + 3}$ and gravitational term $\Omega_G \propto x^{5}$, leading to 
\begin{equation}
      \alpha_{\rm{vir}} \propto x^{2\beta - 2}.
\end{equation}
Therefore, while $\beta < 1$ the virial parameter will decrease as the radius $r$ at which clump properties are evaluated increases.  Most measurements of $\beta$ on $\gtrsim 0.1$ pc scales range between 0.2--0.5 \citep[e.g.,][]{HeyerBrunt2004, caselli_1995, lin2021}, so this condition {\color{black} appears} easily met.  We then expect that, {\color{black} if this condition were met}, a full clump including CO-dark gas would have a lower $\alpha_{\rm{vir}}$ than the result derived only from the CO-traced material, i.e., it would be more gravitationally dominated that could be inferred from CO alone.

\subsection{Plummer Profile}\label{S:plummer}

The Plummer density profile \citep{plummer1911} is frequently applied to molecular clouds and yields a small, flat inner core that transitions to a power-law profile at large radii.  The Plummer profile follows
\begin{equation}
    \rho(r) = 
    \rho_c \left(\frac{1}{\sqrt{x^2+1}}\right)^\eta,
\end{equation}
where {\color{black}$\rho_c$ is the central density}, $R_0$ is the radius of the central core, $x\equiv r/R_0$, and $\eta$ is the index of the power-law at large radii.  \citet{pattle2016} modeled the {\color{black}evolution of pressure-confined cores following Plummer-like density profiles in order to evaluate} whether the cores were likely to collapse or to reach virial equilibrium as a function of radius.  Here we extend this work in the context of CO-dark gas.  

{\color{black}We derive corrections for CO-dark gas for two values of $\eta$: $\eta=2$ as consistent with recent observational results ranging between $\eta=1.5$--2.5 \citep[e.g.,][]{Arzoumanian2011,PalmeirimAndre2013,zucker2021}, and $\eta$=4 following \citet{whitworth2001} and \citet{pattle2016}.} 
We adopt an internal cloud velocity dispersion profile following Equation \ref{eqn:sigmav}.  $\rho(r)$ and $M(r)$ are shown in Figure \ref{fig:profs}  {\color{black}for the fiducial clump with properties } [$R_0=0.1$ pc, $R_{H_2}=1$ pc, $M(R_{H_2})=300 \ M_\odot$ $\sigma_v(R_{H_2})=$ 0.6 km  s\ts{-1}].    

{\color{black}In Appendix \ref{ap:plummer}, we derive the virial parameter for this profile as}
\begin{equation}
    \alpha_{\rm{vir}}(r) =
    \frac{3\sigma_v^2(r)}{\pi \rho_c G R_0^2 } \frac{P_1(r)}{P_2(r)}.
\end{equation}
where
\begin{equation}
P_1(r) =
    \begin{cases}
    {\color{black}4(x - \arctan(x)) \ \rm{for} \ \eta = 2} \\
    {\color{black}2(\arctan(x) - \frac{x}{x^2+1}) \ \rm{for} \ \eta = 4,}
    \end{cases}
\end{equation}
and
\begin{equation}
P_2(r) =
    \begin{cases}
    {\color{black} 4 P_1(r) - 16 \displaystyle\int_0^x \frac{x' \arctan(x')}{x'^2+1} \rm{dx'} \ \rm{for} \ \eta = 2} \\[0.8em] 
    {\color{black}\arctan{(x)} + \frac{x-4\arctan{(x)}}{x^2 +1} + \frac{2x}{(x^2+1)^2} \ \rm{for}   \ \eta = 4.}
    \end{cases}
\end{equation}
{\color{black} We similarly derive}
\begin{equation}
    f_{DG} = 
    \begin{cases} 
    {\color{black}1- \left(\frac{\frac{R_{CO}}{R_0} - \arctan(R_{CO}/R_0) }{ \frac{R_{H_2}}{R_0} - \arctan (R_{H_2}/R_0)}\right) 
    \ \rm{for}   \ \eta = 2} \\[1.5em] 
    
    1 - \left(\frac{  \arctan(R_{CO}/R_0) - \frac{(R_{CO}/R_0)}{(R_{CO}/R_0)^2 + 1} }{ \arctan(R_{H_2}/R_0) - \frac{(R_{H_2}/R_0)}{(R_{H_2}/R_0)^2 + 1} }\right) \ \rm{for}   \ \eta = 4. 
    \end{cases}
    \label{eqn:ap_plummer_fdg}
\end{equation}

In Figure \ref{fig:alphavir}, we present the behavior of \alphavir as a function of $r$ for this profile.    
{\color{black} We also numerically solve for and plot the expected $R_{CO}$ using Equation \ref{eqn:ap_plummer_fdg}} for \fDG = 0.3, 0.5, and 0.8.  For \fDG=0.3, $R_{CO}\simeq$ [{\color{black}0.75 pc for $\eta=2$}, 0.3 pc for $\eta=4$]. For \fDG=0.5, $R_{CO}\simeq$ [{\color{black}0.55 pc for $\eta=2$}, 0.2 pc for $\eta=4$], and for \fDG=0.8, $R_{CO}\simeq$ [{\color{black}0.3 pc for $\eta=2$}, 0.1 pc for $\eta=4$]. 

As in the other profiles considered, the behavior of $\alpha_{\rm{vir}}$ for the Plummer profile is highly variable and is dependent upon the density and velocity assumptions made.  For a cloud with $\beta = 0$, \alphavir is  roughly constant for $x > 1$ and only marginally below the value of \alphavir at $x=1$.  This indicates that the stability that would be inferred from just the CO-traced mass is a fairly accurate representation of the stability of the entire cloud.  In contrast, for $\beta = 0.5$, \alphavir increases rapidly above $R_0$, suggesting that the CO-traced cloud would appear more gravitationally bound than the full cloud at $R_{H_2}$.  

\subsection{Comparison of Density Profiles}\label{S:comp_profs}

{\color{black}We consider the effect of the density profile on \alphavir and the amount by which CO-dark gas changes observed clump properties.  The impact of $k$/$\eta$ on the overall value of \alphavir is similar between all profiles considered, with smaller $k$ leading to higher $\alpha_{\rm{vir}}$ below $R_{H_2}$.  \color{black}In particular, \alphavir is typically $\sim$2x larger for the power-law profiles of \S\ref{S:power} and \S\ref{S:powercore} than in the \alphavir of the $\eta=4$ Plummer profile.}  This is the result of the $\eta=4$ Plummer profile having a much higher proportion of mass centrally concentrated at small $r$ than the $k=1$ and $k=2$ power-law based profiles {\color{black}and $\eta=2$ Plummer profile }(see Figure \ref{fig:profs}).  

Since $\Omega_G \sim G M^2 / r$, concentrating a fixed amount of mass within a smaller area increases the object's gravitational potential and decreases the virial mass.  In contrast, $\Omega_K \sim \sigma_v^2 M$ is not as dependent on the volume in which $M$ is contained so it is unsurprising that \alphavir is significantly reduced for the steeper profiles.  Very subvirial clumps are expected to rapidly collapse, and so to offset this effect and move closer to stability at $R_{H_2}$, the Plummer profile would need to have a much higher $\sigma_v(R_{H_2})$.  
{\color{black}The assumed radial velocity dispersion index $\beta$ also has a large impact on clump dynamical state and the \alphavir that would be inferred after correcting for CO-dark gas}.  For all profiles considered in this section, the choice of $\beta$ generally corresponds to the ``direction'' of the behavior of $\alpha_{\rm{vir}}$ with $r$, whether increasing, decreasing, or constant.  

Throughout this work, we use \fDG for a given clump as a set parameter, without attempting to tie its specific value to the underlying physics that determine the value of \fDG.  In reality, \fDG is a function of the properties of the clump and the environment in which it is immersed.  Since we aim to derive corrections that may be applied by observers using a specific assumed or measured value of \fDG to estimate clump properties, accounting for the nuances of the physical drivers of \fDG is beyond the scope of this work.  {\color{black} However, we do expect that clumps with steeper density profiles will have lower \fDG than clumps with shallower profiles occupying the same environment under identical conditions.  

We can intuitively consider that, in a given environment, a specific $A_V$/density threshold must be reached for CO to effectively self-shield (determined by radiation field strength, dust-to-gas ratio, etc).  Since steeper profiles are more centrally concentrated in mass, they would contain a larger fraction of the total clump mass at this density floor where CO begins to be destroyed ($R_{CO}$); this would decrease $f_{DG}$ despite being in an identical environment to a shallower clump.}

{\color{black}Finally, we evaluate the overall effect of the internal density profile assumed on inferred CO-dark-corrected clump properties.  For the fixed $R_{H_2}$ we consider, the derived $R_{CO}$ for a given \fDG decreases with increasing $k$ (or $\eta$) because of our assumption that $R_0 \ll R_{CO}$.  The values of $R_{CO}$ and $\alpha_{\rm{vir}}(R_{CO})$ for a power-law with core profile with $k=1$ are functionally identical to those for a full power-law, and for a $k=2$ power-law with core profile depart only slightly from the values derived from a full $k=2$ power-law.  Under most scenarios where $R_0 \ll R_{CO}$, the corrected properties ($R_{H_2}$, \alphaHtwo, etc) derived for the power-law with core profile vary by a small amount (generally $\lesssim$10$\%$ difference) from the corrections for a full power-law.  The difference between corrected properties between profiles increases with increasing \fDG. The difference in corrected values between the steep $\eta=4$ Plummer and power-law profiles is larger, but this is likely more of an effect of the variation in assumed $\eta$ vs. $k$ than of profile itself; the $\eta=2$ profile also typically differs by $\lesssim$10$\%$ from the power-law based profiles.  

Therefore, we conclude that the relative steepness by which density decreases with $r$ has a larger impact on the effects of CO-dark gas on observed clump properties than the exact form of the radial density profiles we consider.  For the remainder of this work, we focus our analysis on the behavior of clumps following single power-law profiles for simplicity.}

\section{CO-dark Gas and Size--Linewidth-- Surface Density Relationships}\label{S:effects}

\begin{figure*}
    \centering
    \includegraphics[width=\textwidth]{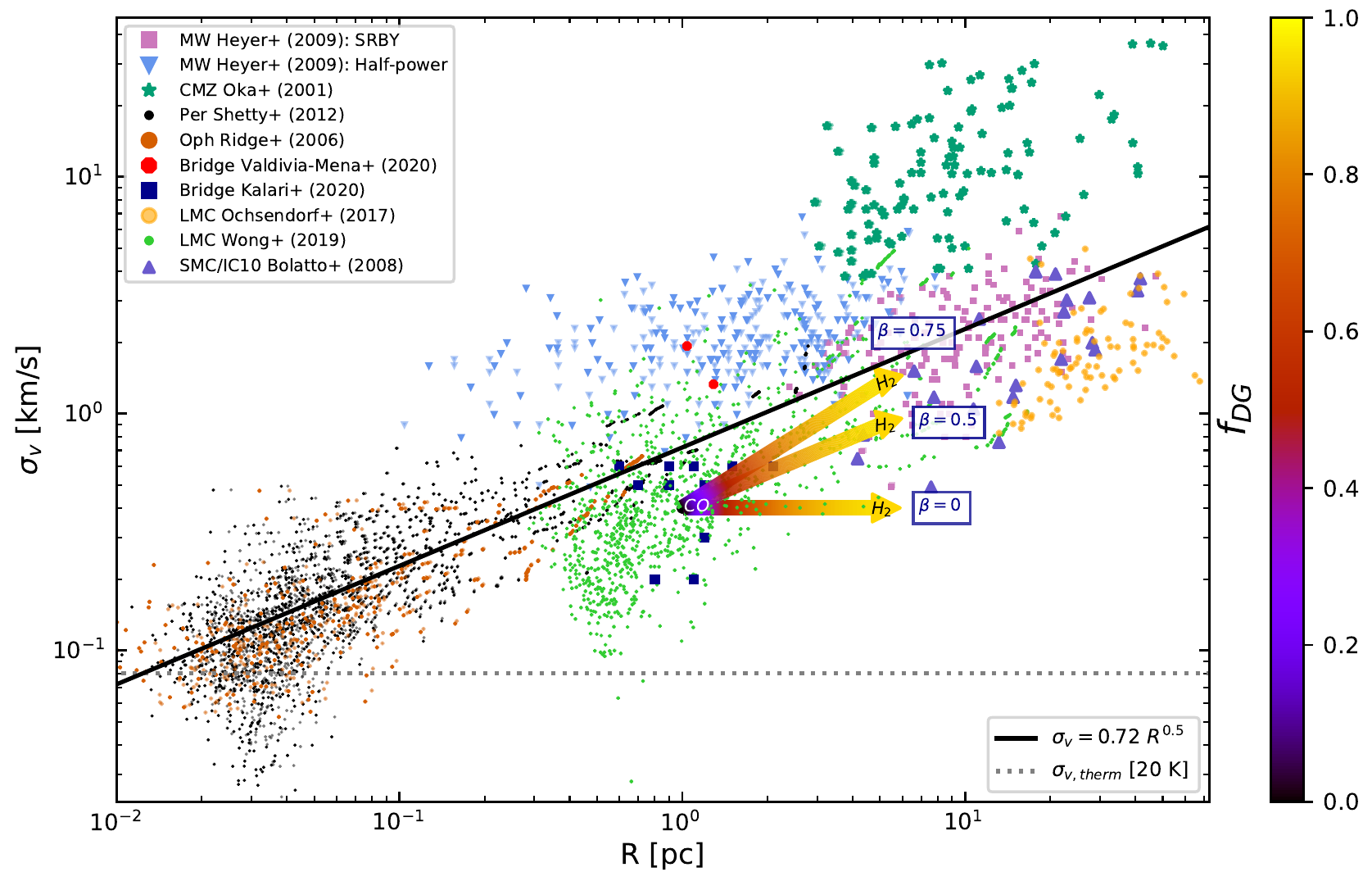}
    \caption{Velocity dispersion $\sigma_v$ compared to radius $R$ of CO-traced structures described in \S\ref{S:effects}.  The black line follows the relationship $\sigma_{v}=0.72 R^{0.5}$ (Equation \ref{eqn:h09_sigR}) and the grey dotted line shows the expected contribution from thermal motion to $\sigma_v$ at T=20 K.  
    The arrows show the direction in which one would correct observed CO-traced clump properties for CO-dark gas to recover the properties of the full H$_2$ clump.  The arrows start at physical properties typical of parsec-scale CO-traced clumps, and move towards the inferred properties of the H$_2$ clump.  Each arrow is labeled with the power-law index $k$ and velocity dispersion index $\beta$ assumed to generate its path, and the color gradient along the arrows shows the corrected H$_2$ properties as a function of \fDG.}
    \label{fig:sizeline}
\end{figure*}

\begin{figure*}
    \centering
    \includegraphics[width=\textwidth]{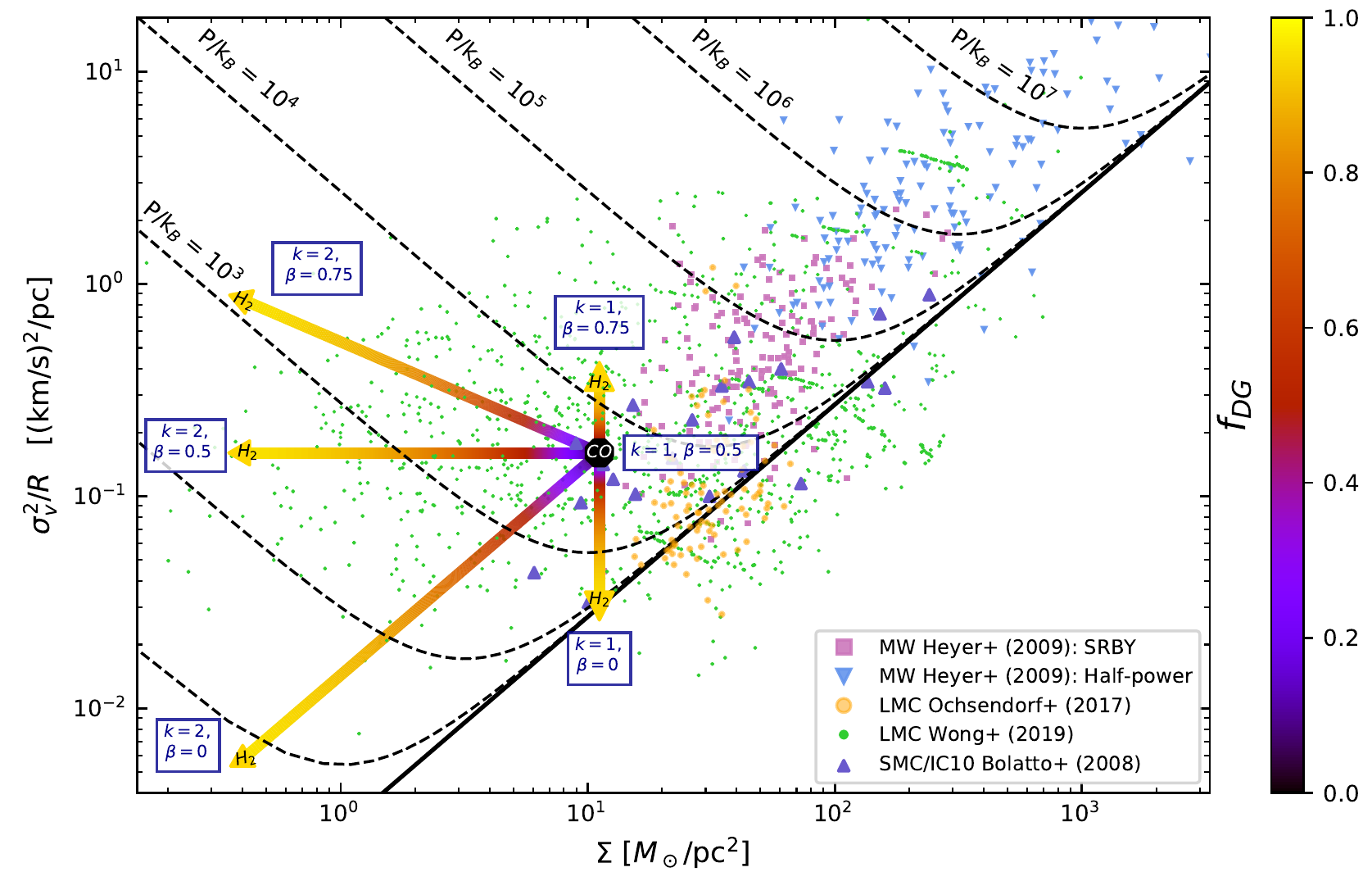}
    \caption{Size-linewidth parameter $\sigma_v^2 / R$ compared to surface density $\Sigma$ for CO-traced structures described in \S\ref{S:effects}.  The black line corresponds to virial equilibrium without external pressure (Equation \ref{eqn:virial_sigma}) and the dashed black curves correspond to virial equilibrium under external pressure with units for P/k$_B$ labels in K cm$^{-3}$ (Equation \ref{eqn:field_pext}).  The arrows are as in Figure \ref{fig:sizeline} and show the direction in which one would correct observed CO-traced clump properties to recover the properties of the full clump including CO-dark H$_2$. }
    \label{fig:sizeline_surfp}
\end{figure*}

\subsection{Larson's Scaling Relationships}

\citet{larson1981} observed correlations between the size $R$, velocity dispersion \sigv, and mass surface density $\Sigma$ of Galactic molecular clouds that have been confirmed and refined by later studies.  The first of these {\color{black}relationships} is a power-law relationship between the size of a cloud $R$ and $\sigma_v$, where 
\begin{equation} 
    \sigma_v \simeq C \left(\frac{R}{1 \ \rm{pc}}\right)^{\Gamma} \rm{km \ s}^{-1}.
    \label{eqn:gen_sizeline}
\end{equation}
\citet{larson1981} originally derived $\Gamma=0.38$ and $C=1.1$ km s\ts{-1}, an estimate that \citet{solomon_mass_1987} and \citet{heyer_re-examining_2009} (hereafter SRBY and H09, respectively) later refined to 
\begin{equation}
    \sigma_v \simeq 0.72 \left(\frac{R}{1 \ \rm{pc}}\right)^{0.5} \rm{km \ s}^{-1}.
    \label{eqn:h09_sigR}
\end{equation}

Larson's second {\color{black}relationship} is derived from observed correlations between $\sigma_v$ and cloud mass $M$, 
\begin{equation}
    \frac{2 \sigma_v^2 R}{G M} \simeq 1,
\end{equation}
which is usually interpreted as meaning that most clouds are roughly in virial equilibrium.  Alternatively, it has been suggested that this is a signature of global hierarchical collapse at all scales within clouds \citep{Ballesteros-Paredes2011,Vazquez-Semadeni2019}.         
Finally, cloud density and size are observed to be inversely related, $n \propto R^{-1.1}$, 
suggesting that surface density is independent of size and should be roughly constant for clouds under conditions similar to the Milky Way, although observations have suggested that $\Sigma$ does vary over several orders of magnitude with environment \citep[e.g., H09;][]{sun2018, traficante2018, DessaugesZavadsky2019, Chevance2020_lifecycle}.

As noted by H09, a natural extension of these {\color{black}relationships is an association} between surface density $\Sigma$ and the size-linewidth parameter $\sigma_v^2 / R$.  A virialized sphere following a power-law density distribution should follow 
\begin{equation}
    \frac{\sigma_v^2}{R} = \frac{(3-k)}{3(5-2k)}  \pi G \Sigma.
    \label{eqn:virial_sigma}
\end{equation}

In Figures \ref{fig:sizeline} and \ref{fig:sizeline_surfp}, we compare the relationships between $R$, $\sigma_v$, and $\Sigma$ for structures observed using CO as a tracer across a variety of environments: 

\begin{enumerate} \itemsep -0.2 em
\item Galactic giant molecular clouds (GMCs), with sub-samples with areas defined from $^{12}$CO by SRBY and from the $^{13}$CO half-power contours of their central cores (H09);
\item clouds in the Galactic central molecular zone \citep{oka_statistical_2001}; 
\item cores observed in the Ophiuchus molecular cloud \citep{ridge_complete_2006}; 
\item cores in the {\color{black}Perseus molecular cloud} \citep{shetty_linewidth-size_2012}; 
\item Clumps in the Magellanic Bridge \citep{kalari_resolved_2020,valdivia-mena_alma_2020}; 
\item Clumps in the LMC regions 30 Doradus, A439, GMC 104, GMC 1, PCC, and N59C \citep{wong_relations_2019}; 
\item GMCs in $\sim$150 star-forming regions throughout the LMC \citep{ochsendorf_what_2017}; 
\item Clouds in the SMC and dwarf galaxy IC 10 \citep{bolatto2008},  
\end{enumerate}
{\color{black} (where the choice of terminology core/clump/cloud corresponds to commonly used size scales of $\sim$0.1/1/10 pc, without any differences in relevant physics implied).}

In Figure \ref{fig:sizeline} where $R$ and \sigv are compared, the usual size-linewidth relationship of Equation \ref{eqn:h09_sigR} is displayed.  $\Sigma$ and $\sigma_v^2/R$ are compared in Figure \ref{fig:sizeline_surfp} for a subset of the sources listed above that have cloud mass estimates derived without assuming virial equilibrium (1, 6, 7, 8 in the list above).  Equation \ref{eqn:virial_sigma} is shown as the straight black line for $k=0$.  Additionally, in pressure-bounded virial equilibrium, $\Sigma$ and $\sigma_v^2 / R$ are related as \citep{field_does_2011} 
\begin{equation}
    \frac{\sigma_v^2}{R} =  \frac{1}{3} \left(\pi \gamma  G \Sigma + \frac{ 4 P_{e}}{ \Sigma}\right),
    \label{eqn:field_pext}
\end{equation}
which is shown in Figure \ref{fig:sizeline_surfp} by the V-shaped curves, with $\gamma=0.6$ for a cloud with $k=0$. 

The majority of the \citet{wong_relations_2019} and \citet{ochsendorf_what_2017} 
LMC GMCs, \citet{kalari_resolved_2020} Bridge clumps, and \citet{bolatto2008} SMC and IC 10 clouds have \textbf{smaller \sigv for a given $R$ than expected from Galactic clouds}, falling well under the relationship described in Equation \ref{eqn:h09_sigR}.  This has been observed in a variety of other low-Z environments as well,  e.g. by \citet{Rubio2015} in the $Z \simeq 0.13  \ Z_\odot$ dwarf galaxy Wolf–Lundmark–Melotte, and \citet{Hughes2013} in the LMC.  Many of these samples also have lower $\Sigma$ for a given $\sigma_v^2 /R$ than expected based on Equation \ref{eqn:virial_sigma}, suggesting that the structures are either unbound and transient or must be confined by external pressure to remain stable, as position in this space is directly related to $\alpha_{\rm{vir}}$.

{\color{black}As part of the PHANGS-ALMA collaboration \citep[“Physics at High Angular-resolution in Nearby GalaxieS with ALMA,” ][]{leroy2021_phangs}, \citet{SunLeroy2020_dynamical} analyzed the dynamical states of molecular gas in 28 nearby disk galaxies.  They derived typical midplane pressures over 1 kpc scales ranging from $P/k_B = 10^3$--10$^6$ K cm$^{-3}$, and found that the average internal turbulent pressure of clouds was typically very similar to the required cloud-scale equilibrium pressure, which they concluded indicated that most gas was in dynamical equilibrium.}  
  {\color{black} \citet{WongHughes2009} derived an average midplane hydrostatic pressure in the central regions of the LMC of $P/k_B \sim 10^4$ K cm$^{-3}$ using \HI{} and CO(1--0) observations, which could be sufficient to confine a large fraction of the \citet{wong_relations_2019} LMC clumps observed to have high $\sigma_v^2 /R$, as well as the majority of the \citet{ochsendorf_what_2017} LMC GMCs.}

\subsection{Effects of CO-dark Gas on Observed Relationships}

The effects of our derived corrections for CO-dark gas in a power-law density profile clump (\S\ref{S:power}: Equations \ref{eqn:rh2_plaw}, \ref{eqn:fdg_sigv}, and \ref{eqn:fdg_surfp}) are shown by the arrows in Figures \ref{fig:sizeline} and \ref{fig:sizeline_surfp}.  The arrows start at the properties of a clump as observed solely in CO and move towards the ``true'' characteristics of the full clump including CO-dark gas, with color gradients along the arrow corresponding to \fDG.  The initial conditions for the corrections displayed are [$R_{CO}=1$ pc, $\sigma_{v}(R_{CO})=0.4$ km s\ts{-1}, $M(R_{CO})=35 \ M_\odot$]; these values correspond to the medians of these quantities for roughly pc-scale CO clumps in the \citet{wong_relations_2019} sample.  Changing the {\color{black} arrow's origin does not impact the direction of the arrow.}  In Figure \ref{fig:sizeline}a, we only show corrections for $k=1$, with arrows for $\beta = 0$, $\beta = 0.5$, and $\beta=0.75$; this is because the corrections for $k=1$ vs $k=2$ overlap in this space and differ only in the extent to which their arrows extend.  In Figure \ref{fig:sizeline_surfp} we also show corrections for $k=2$.

For $\beta = 0$, correcting for CO-dark gas causes clumps to have even lower \sigv relative to the increased $R$ and thus drives the clumps further from following Equation \ref{eqn:h09_sigR}.  We note that a velocity profile this ``flat'' is unlikely, as turbulence within the ISM is mainly driven at large scales \citep{brunt2009}, but we display it to demonstrate the limits of this effect.  For $\beta=0.5$, the corrections have no effect on the position of the clump in size-linewidth space relative to the expected Equation \ref{eqn:h09_sigR} relationship; this is because the standard inter-clump relationship Equation \ref{eqn:h09_sigR} and the displayed intra-clump profile share the same $\beta = \Gamma = 0.5$.  By the same logic,  $\beta=0.75$ unsurprisingly brings clumps closer to agreement with Equation \ref{eqn:h09_sigR} because $\beta > \Gamma$.   
Corrections in $\Sigma$ vs. $\sigma_v^2/R$ space have a similarly variable effect.  The distance of any given clump from the virial line in Figure \ref{fig:sizeline_surfp} is directly proportional to the stability of the clump as measured by \alphavir and we interpret the corrections {\color{black} for a power-law profile} in this context as follows.  
\begin{itemize}
\item We again see that the assumed $k$ and $\beta$ have a large impact on the inferred corrected state: for $k=1$, $\Sigma$ is constant; while for $k>2$, the corrected $\Sigma$ is significantly reduced.
  
\item We observe that clumps decrease in \alphavir and move towards \alphavir$\sim 1$ in all cases where $k+2\beta < 2$.  This suggests that, if these profile conditions are met, the apparently high \alphavir structures traced by CO in low-Z, high \fDG environments may be closer to stability than expected.  

\item In most cases, the updated clump positions suggest that a lower level of external pressure would be required to maintain stability than would be inferred from CO-traced material alone.  
\end{itemize}

The assumed density and velocity profiles then almost entirely determine the ``direction'' of these biases.  This highlights the importance of studies of the spatial dependence of density and linewidth on the scale of individual clouds in addition to SRBY/H09-type studies comparing these quantities between cloud populations.

\section{Discussion}\label{S:discuss}

\subsection{Can CO-dark Gas Explain Departures from Larson's Relationships?}\label{S:explain}

From the clump property corrections derived in \S\ref{S:density} and described in \S\ref{S:effects}, it is clear that neglecting CO-dark gas could significantly bias the assessment of cloud placement in Larson's {\color{black}relationships} and gravitational stability.  We now examine if this effect is sufficient to explain the observed high $\alpha_{\rm{vir}}$ and departures from Larson's relationships in low-Z environments.

Under the corrections for a power-law profile that we have derived, low-$\sigma_v$ clumps must follow an internal velocity profile with $\beta > 0.5$ (i.e., have large motions at large scales) to reconcile with the typical size-linewidth relationship described by Equation \ref{eqn:h09_sigR}; however, large $\beta$s also yield increased \alphavir that imply the full structure is gravitationally unbound.  If instead one assumes that clouds are close to virialized without external pressure, then the dark gas correction required to move observed points closer to virialization (i.e., to decrease \alphavir) requires that clumps follow a shallow density profile and have $\beta < 0.5$ --- but, shallow $\beta$s increase the amount by which these clouds fall ``under'' the $R$--$\sigma_v$ relationship of Equation \ref{eqn:h09_sigR}. 

This contradiction is most problematic in structures with high \fDG as expected in low-Z or high-radiation environments, and can be resolved if clouds in these areas are: (1) overwhelmingly gravitationally unstable and dispersing rapidly as a result; or (2) require much higher levels of external pressure to remain stable than clouds in more typical environments; or (3) possess a global \sigv/R trend shifted to lower values of \sigv than the classical Equation \ref{eqn:h09_sigR} relationship (i.e., a smaller scaling coefficient $C$ in Equation. \ref{eqn:gen_sizeline}) and have shallow internal density and \sigv profiles (0$\leq\beta<$0.5).  

(1) is unlikely statistically simply because of the number of clouds that are observed, and a physical cause for (2) is hard to imagine since the typical ISM pressure in low-Z galaxies is $\sim$1--2 orders of magnitude smaller than in typical large spiral galaxies \citep{de_los_reyes_2019}.  There are also nontrivial direct relationships between metallicity and ISM pressure in these areas because of reduced cooling and thermal balance, but predictions as a function of metallicity are generally only possible in the context of a self-regulated star formation model and thus the specifics depend on the details of that model.  Additionally, the direct effects of metallicity via the cooling rate on pressure are less important than the galaxy type to the properties of molecular clouds

(3) is then the most compelling, and would be the simplest way to account for observed low-\sigv and high \alphavir structures in low-Z areas.  Shallow density profiles of $1.5 < k < 2$ are typical on the pc-scales where the simplified isolated spherical PDR model that we consider here holds \citep{caselli2002,pirogov09, Arzoumanian2011, Schneider2013}, and even shallower profiles ($k\sim1$) have been found in young, low-density cores and clumps \citep{chen2019,chen2020,lin2021}.  
{\color{black} Small values of $C$ and steep $\Gamma$ relative to SRBY's $C=0.72$ and $\Gamma=0.5$ have been derived from CO observations} for structures in the SMC, LMC, and other local dwarf galaxies where low-\sigv / high \alphavir structures are found (with C $\sim$ 0.2 -- 0.6 and $\Gamma \sim$ 0.55--0.85) 
\citep{bolatto2008, hughes2010_magma, Hughes2013, wong_relations_2019}.  {\color{black} In CO-dark regions, \HI{} can also be used as a probe of turbulence.  For a sample of \HI{} clouds in the LMC, \citet{KimRosolowsky2007} derived a mean $\Gamma\simeq0.5$.}

In the pioneering \citet{larson1981} study, a shallow $\Gamma = 0.38$ was derived, which is similar to the Kolmogorov index for turbulent cascade in an incompressible medium $\beta \sim 1/3$.  More recently, $\beta \simeq 1/2$ has frequently been found for GMCs both observationally and through simulations \citep[e.g.,][]{HeyerBrunt2004,Dobbs2015}; this aligns the expectation for Burgers turbulence \citep{Passot1988} i.e., in an isotropic system dominated by shocks, and is in accordance with SRBY's $\Gamma=0.5$. On very small scales ($\lesssim 0.05$ pc) a break in the internal size-linewidth relationship has been observed with $\beta$ approaching zero \citep{goodman_coherence_1998, caselli2002, volgenau2006, Pineda2010}; however, it seems unlikely that the W10 scenario of PDRs of isolated individual spherical clouds surrounded by envelopes of dark gas would hold on these sizes because cores are typically embedded within larger structures.  

Shallow values of $\beta$ ($\beta \sim$ 0.2--0.3) have also been derived in high mass star-forming regions \citep{caselli_1995} and in prestellar cores and young clumps \citep{Tatematsu2004, lee2015_sfr,lin2021}.  \citet{Bertram2015_delta} analyzed turbulence within simulated molecular clouds using the $\Delta$-variance method, from which they compared the values of $\beta$ within the full cloud, within H$_2$ gas, and within CO-traced material.  For initial densities ranging between 30 -- 100 cm$^{-3}$, the derived $\beta$ ranged between $\sim$0.3--0.6 as derived from the resulting H$_2$ density maps, and $\sim$0.15--0.4 as traced by CO density, a difference which they attributed to the compact nature of the CO structures as compared to the more extended H$_2$.  

We emphasize that the inter-clump size-linewidth relationship with exponent $\Gamma$ is obtained by comparing populations of clumps, while the intra-clump size-linewidth relationship with exponent $\beta$ is obtained by studying individual structures.  The latter relationship is much more challenging to measure in the typically distant low-Z environments due to the required high angular resolutions and has only recently become possible, but is key for assessing if the implied shallow $\beta$ is realistic.  Overall, the measurements of $\beta$ that have been obtained locally generally resemble observed values of $\Gamma$.  

{\color{black} This observed correspondence of $\beta \sim \Gamma \simeq 0.5$ has been interpreted as reflecting the uniformity of velocity structure functions between individual clouds, so that $\Gamma$ is largely set by $\beta$ \citep{HeyerBrunt2004}.   The implication from (3) that $\beta$ is shallower than the observed $\Gamma$ in low-Z environments creates some tension with this conclusion.  One explanation for this difference could be a correlation between $f_{DG}$ and cloud size.  In their sample of LMC GMCs, \citet{ochsendorf_what_2017} observed a decrease in the ratio of CO-traced mass to dust-traced mass as dust-traced mass increased.  Since the dust-traced mass likely includes the diffuse CO-dark gas, this suggests that a correlation between \fDG and cloud size exists with larger clouds having higher \fDG.  Larger clouds would then systematically have larger relative changes between their true properties including CO-dark gas to their observed properties than smaller clouds do.  

The ``true'' $\Gamma$ relating the full clouds including CO-dark gas could then be shallower than the observed, CO-derived $\Gamma$, and instead approach (and possibly be determined by) the expected shallow $\beta$.  This would explain the general steepness of CO-traced $\Gamma$ in low-Z environments, as well as resolve the implied difference between $\beta$ and $\Gamma$ in low-Z environments.}  It is of course also possible that clumps in these low-Z environments do truly have different physical properties and scaling relationships than clumps under Galactic conditions.

\subsection{CO-dark Gas and Star Formation}

\subsubsection{Star Formation Efficiency Considering CO-dark Gas}

On kpc-scales, low-Z galaxies have been found to depart from the Kennicutt-Schmidt relationship, possessing higher star formation rate densities at a given molecular gas surface density as assessed by CO than found in more typical environments \citep[e.g.][]{Galametz2009,Schruba2012}.  Star formation efficiency (SFE) is frequently assessed by comparing the star formation rate (SFR) to gas mass  ($\epsilon'=$ SFR / $M_{\rm{cloud}}$), so this departure suggests that the SFE is also much higher than under Galactic conditions.   

\citet{madden2020} showed that CO-dark gas is sufficient to cause the apparent variation from the Kennicutt-Schmidt relationship on galactic scales, and that when corrected for the missing mass star formation in these environments is not significantly more efficient.  It has also been suggested that H$_2$ gas is not a requirement for star formation but is usually present as a consequence of the necessary shielding for stars to form \citep{Glover2012_atomic, Krumholz2012_atomic}.  Star formation could then in principle proceed in atomic gas without the presence of molecular gas (although this would be rare), and may explain the lack of CO detections and corresponding high implied SFEs in some low-Z star-forming galaxies.

While CO-dark gas appears to be responsible for increased SFEs on large scales because surface densities averaged over large scales are increased by the addition of CO-dark gas mass, it is unclear how it impacts star formation in individual clumps.  SFE is also frequently evaluated by simply comparing the total stellar mass to total molecular mass ($\epsilon = M_{*}/M_{\rm{cloud}}$), or as a function of free-fall time ($\epsilon_{ff}$=$\tau_{ff} \times \epsilon'$).  A simple but perhaps naive correction to the SFE of an individual CO-traced clump for missing H$_2$ mass would be $\epsilon_{H_2} = (1-f_{DG}) \epsilon_{CO}$ by Equation \ref{eqn:mDG}, {\color{black}with a similar correction for $\epsilon'$.  We have shown that CO-based observations are likely to overestimate mean clump density. This would lead to underestimates of free-fall time $\tau_{ff}$ and also, depending on density profile, potentially to underestimates of $\epsilon_{ff}$ as well.   }

However, the $\epsilon$-based metrics are generally derived over larger scales, which helps offset the unknown variation from the original total gas mass for any given star-forming clump to its present day mass by averaging over clumps and cores at a variety of stages in the star formation process.  Using the present-day gas mass of a single clump to try to derive a by-clump efficiency loses this advantage, and so we only suggest the use of the proposed corrected $\epsilon_{H_2}$ and related quantities on larger scales and even then with caution.  

It is still not well understood if the actual way and timescale over which clouds collapse in low-Z environments is different than under conditions similar to the solar neighborhood, and, if so, how this departure influences the SFR/SFE.  \citet{parmentier2020} and \citet{parmentier_pasquali_2020} derived a relationship between clump radial density profile and SFR and found that steeper profiles correspond to higher initial SFRs: star formation proceeds most rapidly in the densest areas of clumps, and the centers of clumps with very steep density profiles are denser than shallower clumps of the same mass.  

Since radiation fields are known to be enhanced in the interclump medium due to the decreased dust-to-gas ratios, it is plausible that the typical radial profiles of clumps could be different than in higher-Z environments. We have shown that shallow density profiles are required for the properties of low-Z clumps to approximate those of Galactic clumps, so it follows that in this scenario the low-Z clump-scale SFR could be slower than in higher-Z environments.  To reconcile with the observed high SFR averaged over kpc-scales, relatively more clumps would need to exist to achieve these values.   

{\color{black} Measurements of the total gas mass and SFR over large scales is clearly critical for these observations of SFE, and it is well understood that underestimating total mass can skew SFE estimates.  An additional factor is the mechanics of how these clouds collapse to form stars at clump and core scales and the fraction of gas at these scales that actively contributes to star formation. The extent to which the diffuse CO-dark envelopes participate in star formation is unclear and is one of several contributing factors that sets SFE.} 

{\color{black} The scaling relationships between molecular gas and SFR observed over large scales can be validated by understanding the fraction of gas at clump-scales involved in star formation and the factors that affect the stability of individual cores and clumps.  Detailed studies of the distribution and state of clumpy molecular gas is key in fully explaining the SFR/SFE and origin of scaling relationships at kpc-scales.  Our models show the importance of CO-dark gas fraction and density and velocity dispersion profiles in influencing these properties.  This work is then relevant to large scale measurements of SFEs in contextualizing interpretations of these measurements.  }

\subsubsection{Relationship between $f_{DG}$ and the CO-to-H$_2$ Conversion Factor}

The corrections for CO-dark gas we have derived are dependent on having an estimate of the total molecular gas within $R_{CO}$, which could be derived through e.g., assuming local thermal equilibrium (LTE) with the use of multiple CO transitions, applying the non-LTE RADEX modeling, or similar methods.  The widely-used CO-to-\Htwo conversion factor
\begin{equation}
\begin{split}
    X_{CO} &= \frac{\rm{N}(\rm{H}_2)}{I_{CO}} \ [\rm{cm}^{-2} \ (\rm{K \ km \ s}^{-1})^{-1}] \\ 
    \alpha_{CO} & = \frac{M_{\rm{vir}}(R_{CO})}{L_{CO}} \ [M_\odot \ (\rm{K \ km \ s}^{-1} \ \rm{pc}^2)^{-1} ],
\end{split}
\end{equation} 
is also designed to account for the untraced H$_2$ gas that we correct for in this work using \fDG.  Some degree of correspondence between the two is then expected, as shown in previous works simulating the relationship between $X_{CO}$ and environmental conditions \citep{Shetty2011_XCO_1,Shetty2011_XCO_2,Clark2015_xco,szucs_how_2016,gong2018,gong_environmental_2020}.  

To demonstrate this expected relationship in the context of this work, we define a crude mass ratio factor $Y_{DG}$ where
\begin{equation}
    Y_{DG} = \frac{M(R_{H_2})}{M(R_{CO})} = \frac{1}{1-f_{DG}}, 
\end{equation} 
by Equation \ref{eqn:fdg_Gen}.  To compare the value of $Y_{DG}$ across different environments, we define $Y_{DG,MW}$ as the typical value of $Y_{DG}$ in the $f_{DG,MW}$ Milky Way.  The expected $Y_{DG}$ in a given environment can then be compared to $Y_{DG,MW}$ through the ratio of their respective $f_{DG}$,
\begin{equation}
    Y_{DG} = \left(\frac{1-f_{DG,MW}}{1-f_{DG}}\right) \ Y_{DG,MW}.
    \label{eqn:ydg}
\end{equation}

For a typical $Z \sim 1 \ Z_\odot$ Galactic environment with $f_{DG,MW}=0.3$ (W10), $Y_{DG,MW}\simeq 1.4$.  From this value, $Y_{DG} \simeq 3.5 \ Y_{DG,MW}$ would be expected in an environment like the  $Z=0.2 \ Z_\odot$ SMC with \fDG $\sim$ 0.8 \citep{jameson_first_2018}.  This corresponds very well to the observed ratio between the usual Galactic $X_{CO,MW}$ and $X_{CO}$ derived in SMC clumps: for sub-pc clumps in the SMC Wing \citet{muraoka_alma_2017} derived $X_{\textrm{CO}}\sim$ 4 $X_{\textrm{CO,MW}}$, and for pc-scale clumps in the Magellanic Bridge \citet{kalari_resolved_2020} derived $X_{\rm{CO}}\sim$3 $X_{\textrm{CO,MW}}$ and \citet{valdivia-mena_alma_2020} found $X_{\rm{CO}}\sim$1.5--3.5 $X_{\textrm{CO,MW}}$.  This suggests that $Y_{DG}$ (and \fDG) could be used as a check of measured $X_{CO}$ in clumps, or vice versa.  In contrast, measurements of $X_{CO}$ over cloud scales and larger ($\gtrsim$10 pc) in the SMC have ranged between 20--50 $X_{\textrm{CO,MW}}$ \citep{leroy_structure_2009,bolatto_state_2011,
Jameson2016}, significantly exceeding the ratio between $Y_{DG}$ and $Y_{DG, MW}$ that we have derived here.  

This variation between $X_{CO}$ and $Y_{DG}$ estimates is likely caused by the well-known limits of $X_{CO}$ at small scales \citep[see][for a review]{bolatto_co--h2_2013} and the limits of \fDG as formulated for isolated spherical PDRs by W10/in this work at large scales.  When at low metallicities, $X_{CO}$ is expected and observed to increase rapidly.  On cloud and global scales, $X_{CO}$ measurements are averaged over many clouds and so include both diffuse and dense molecular gas.  For individual low-Z/high-\fDG clumps, though, only dense gas is reflected.  

The scale (and resolution) at which clumps are measured is negatively associated with derived $X_{CO}$: in the LMC, for example, \citet{fukui2008_NANTEN} derived $X_{CO} \sim 4 \ X_{CO,MW}$ from clouds observed at $\sim$40 pc resolution by NANTEN, while \citet{hughes2010_magma} derived $X_{CO}\sim \ 2 \ X_{CO,MW}$ from structures observed at $\sim$10 pc resolution by the Magellanic Mopra Assessment (MAGMA) survey.  Lower-resolution observations run the risk of small clumps being diluted by large beam sizes, artificially inflating \alphavir and $X_{CO}$ and also increasing the likelihood of such clumps not being identified at all.  Resolved observations of individual pc-scale clumps in distant low-Z environments have only recently become possible and typically yield smaller conversion factors \citep{muraoka_alma_2017, schruba_physical_2017, saldano_molecular_2018, kalari_resolved_2020, valdivia-mena_alma_2020}, approaching $X_{CO,MW}$ and in alignment with our expectations for clump \fDG.

\subsection{Guidance for Interpreting Observations}\label{S:observers}

We present the case of a ``typical'' observed CO clump with high $\alpha_{\rm{vir}}$, and discuss the properties that would be inferred by an observer using our derived corrections for CO-dark gas.  For a clump following a typical $k=1.5$ and $\beta=0.5$ with [$R_{CO}=1$ pc, $\sigma_{v}(R_{CO}) = 0.4$ km s\ts{-1}, $M(R_{CO})=35 \ M_\odot$, \alphaco$=4.3$], a moderate Galactic $f_{DG}\sim 0.3$ (W10) yields a relatively small difference in clump properties: [$R_{H_2}=1.3$ pc, $\sigma_{v}(R_{H_2})=0.45$ km s\ts{-1}, $M(R_{H_2})=50 \ M_\odot$, \alphaHtwo$=4.8$].  In contrast, an extreme $f_{DG} \sim 0.9$ as occasionally derived in low-Z environments \citep{jameson_first_2018} would lead to a significantly different set of inferred properties: [$R_{H_2}=4.6$ pc, $\sigma_{v}(R_{H_2})=0.86$ km s$^{-1}$, $M(R_{H_2})=350 \ M_\odot$, \alphaHtwo$=9.2$].  

We then see that under typical assumed clump density and velocity profiles, correcting for CO-dark gas does \textit{not} resolve the apparent instability of the structure --- it actually exacerbates the issue.  We emphasize again that the changes in clump properties post-correction are highly dependent upon the choice of $k$ and $\beta$; if the same clump followed shallower profiles, a reduction in $\alpha_{\rm{vir}}$ could just as easily be indicated.  At the same time, the magnitude of this shift makes clear that correcting for CO-dark gas is essential for an accurate assessment of clump properties in high \fDG environments. 

\section{Conclusions}\label{S:conclude}

We have derived easily-applied corrections to CO-derived clump properties to account for the effects of CO-dark gas.  Our main conclusions are as follows:

\begin{enumerate}
    \item For molecular clouds following power-law or Plummer density profiles, CO-derived measurements will systematically underestimate cloud mass and size.  If clumps have shallow mass density and radial velocity dispersion profiles, the virial parameter \alphavir will be overestimated {\color{black}(\S\ref{S:density})}.
    
    \item In order to interpret CO observations as accurately as possible, cloud properties (e.g., size, mass, surface density, velocity dispersion, virial parameter) should be corrected using the prescriptions outlined in \S\ref{S:density} as demonstrated in \S\ref{S:observers}.
    
    \item CO-derived measurements are most suspect in low-Z, high \fDG regions; however, CO-dark gas is unlikely to simultaneously be the cause of observed clumps with high $\alpha_{\rm{vir}}$ and low $\sigma_v$ 
    relative to Larson's relationships in low-Z environments {\color{black}(\S\ref{S:effects}, \S\ref{S:explain})}.
\end{enumerate}

Understanding what other processes might drive departures from Larson's {\color{black}relationships} and from inferred virial equilibrium should be of high priority.  Attempts to correct for all of the above effects are reliant on accurate assessment of intra-cloud density and velocity profiles, and so this too should continue to be prioritized, especially on clump scales.    

It is clear that assessing how star formation proceeds within clumps in low-Z regions is dependent on understanding the impact of CO-dark gas.  Accounting for CO-dark gas both on local and global scales is then key in evaluating the evolutionary history and likely future of specific regions and in placing star formation in low-Z environments into its correct context.  The corrections we have presented here are one tool to better leverage CO observations to estimate clump behavior after accounting for the effects of environment.    

\begin{acknowledgements}
We thank Kelsey Johnson, Allison Costa, and Molly Finn for useful comments and discussion.  T.J.O. and R.I. were supported during this work by NSF award 2009624.  The National Radio Astronomy Observatory is a facility of the National Science Foundation operated under cooperative agreement by Associated Universities, Inc.  This paper makes use of the following ALMA data: ADS/JAO.ALMA\#2012.1.00683.S,  ADS/JAO.ALMA\#2015.1.1013.S,  ADS/JAO.ALMA \#2016.1.00193.S.  ALMA is a partnership of ESO (representing its member states), NSF (USA) and NINS (Japan), together with NRC (Canada), MOST and ASIAA (Taiwan), and KASI (Republic of Korea), in cooperation with the Republic of Chile. The Joint ALMA Observatory is operated by ESO, AUI/NRAO and NAOJ.

\end{acknowledgements}

\software{Astropy \citep{astropy_2013,astropy_2018}; BioRender (\url{https://biorender.com}); Matplotlib \citep{matplotlib_Hunter2007}; Numpy \citep{harris2020_numpy}}

\appendix
\section{Derivation of Clump Density Profiles}\label{ap:profs}
\vspace{1.5em}
For a clump following a given density profile $\rho(r)$, the relevant terms for {\color{black} calculating} the virial parameter and corrections for CO-dark gas that we derive are as follows.  Mass {\color{black} within a radius $r$} can be found as 
\begin{equation}
    M(r)=4\pi \int_0^r  r'^2 \rho(r')  dr'.
    \label{eqn:Mr_gen}
\end{equation}  
This leads to gravitational potential energy
\begin{equation}
   \Omega_G(r) = - 4 \pi G \int_0^r   r' \ \rho(r') M(r') dr',
   \label{eqn:omegaG}
\end{equation}
where $G$ is the fundamental gravitational constant.  
We assume a one-dimensional radial velocity dispersion profile $\sigma_v(r)$ following Equation \ref{eqn:sigmav}.  The total kinetic energy can be found as
\begin{equation}
    \Omega_K(r) = \frac{3}{2} M(r) \sigma_v^2(r),
    \label{eqn:omegaK}
\end{equation}
and the virial mass $M_{\rm{vir}}(r)$ follows from requiring $2\Omega_K(r) = -\Omega_G(r)$.   
\vspace{1.5em}
\subsection{Power-law Profile}\label{ap:power}

We consider a clump following a power-law profile, 
\begin{equation}
    \rho(r) = \rho_c \ x^{-k},
\end{equation}
where $\rho_c$ is the central density, $x=r/R_0$, and $R_0$ is an arbitrary radius at which $\rho$ is normalized.  From Equation \ref{eqn:Mr_gen}, the mass of such a structure is 
\begin{equation}
\begin{split}
    M(r) & =  \pi \rho_c R_0^3 T_1(r),
\end{split}
\end{equation}
where $T_1(r) = \left[\frac{4x^{(3-k)}}{(3-k)}\right]$.    
From Equation \ref{eqn:omegaG}, the gravitational term is
\begin{equation}
\begin{split}
    \Omega_G (r) 
     &= -\pi^2 \rho_c^2 G R_0^5 T_2(r) 
      = -\frac{G M(r)^2}{R_0} \frac{T_2(r)}{T_1(r)^2},
\end{split}
\end{equation}
where $T_2(r) = \left[\frac{16 x^{(5-2k)}}{(5-2k)(3-k)}\right]$. (The use of $T_1(r)$ and $T_2(r)$ in these expressions will make the parallels with subsequent radial density profiles clearer.)  
From Equation \ref{eqn:omegaK}, the kinetic term is 
\begin{equation}
    \Omega_K(r) = 
    \frac{3}{2} \pi \rho_c R_0^{3} T_1(r) \sigma_v^2(r).
\end{equation}
Requiring $2\Omega_K = -\Omega_G$, we derive the virial mass
\begin{equation}
    M_{\rm{vir}}(r) = \frac{3 \sigma_v^2(r) R_0}{G} \frac{T_1(r)^2}{T_2(r)},
\label{eqn:plaw_mvir}
\end{equation}
which is equivalent to the classical virial mass definition  \citep{solomon_mass_1987,MacLaren1988}
\begin{equation}
    M_{\rm{vir}}(r) = \frac{3(5-2 k)}{(3-k)} \  \frac{r \ \sigma_{v}^2(r)}{G}.
\end{equation} 
{\color{black} From Equation \ref{eqn:alvir},} the virial parameter is then
\begin{equation}
\begin{split}
    \alpha_{\rm{vir}}(r) & = \frac{3 \sigma_v^2(r)}{\pi \rho_c G R_0^2} \frac{T_1(r)}{T_2(r)},
\end{split}
\end{equation}
which is equivalent to
\begin{equation}
    \alpha_{\rm{vir}}(r) = \frac{3 (5-2k)}{4 \pi \rho_c} \frac{ \sigma_v^2(R_0)}{G} x^{2\beta+k-2},
    \label{eqn:alpha_of_x}
\end{equation}
under the scaling of the velocity dispersion profile of Equation \ref{eqn:sigmav}.  

{\color{black}In their Appendix A, W10 derived the dark-gas fraction for a power-law density profile with $k<3$ as} 
\begin{equation}
f_{DG}  = 1 - \left( \frac{R_{CO}}{R_{H_{2}}} \right)^{3-k}. 
\end{equation}
{\color{black}$R_{H_2}$ can then be solved for analytically, with $\Sigma(R_{H_2})$, $\alpha_{\rm{vir}}(R_{H_2})$, and related properties following as demonstrated in \S\ref{S:power}.}

\vspace{1.5em}
\subsection{Power-law Profile with Constant Density Core}\label{ap:core}

We consider a clump following a power-law density profile with a uniform core of radius $R_0$,
\begin{equation}
\rho(r) = \begin{cases} \rho_c \ \rm{for} \ r < R_0 \\ 
\rho_c \ x^{-k} \ \rm{for} \ r \geq R_0,
\end{cases}
\end{equation}
where $\rho_c$ is the central density and $x = r/R_0$.  {\color{black}From Equation \ref{eqn:Mr_gen},} the mass within $r$ is
\begin{equation}
    M(r) = \begin{cases} 
    \pi\rho_c R_0^3 \left[\frac{4}{3} x^3\right] \ \rm{for} \ r < R_0 \\
     \pi\rho_c R_0^3 \Pi_1(r) \ \rm{for} \ r \geq R_0,
    \end{cases} 
\end{equation}
where $\Pi_1(r) = \left[ \frac{4}{3-k} \left(x^{3-k} -\frac{k}{3}\right) \right]$.  {\color{black}From Equation \ref{eqn:omegaG},} the gravitational term is
\begin{equation}
    \Omega_G(r)  \equiv \begin{cases}
    -\frac{16}{15} \pi^2 G \rho_c^2 R_0^5 x^5 \ \rm{for} \ r < R_0 \\ 
    -\pi^2G\rho_c^2 R_0^5 \Pi_2(r) \  \rm{for} \ r \geq R_0,
    \end{cases}
\end{equation}
where 

\begin{equation}
    \Pi_2(r) = 
    \begin{cases}
    {\color{black}\frac{16}{3-k} \left(\frac{x^{5-2k}-1}{5-2k} + \frac{k(1-x^{2-k})}{6-3k} + \frac{3-k}{15} \right) \  \rm{for} \ k \neq 2} \\[0.5em]

    {\color{black}\frac{16}{3-k} \left( \frac{x^{5-2k}-1}{5-2k} - \frac{k \ln(x)}{3} + \frac{3-k}{15}    \right) \  \rm{for} \ k = 2. }
    \end{cases}
\end{equation}

{\color{black}From Equation \ref{eqn:omegaK},} the kinetic term is 
\begin{equation}
    \Omega_K(r) = \begin{cases}
    2 \pi \rho_c R_0^3 x^3 \sigma_v^2(r)  \ \rm{for} \ r < R_0 \\
    \frac{3}{2} \pi \rho_c R_0^{3} \Pi_1(r) \sigma_v^2(r) \  \rm{for} \ r \geq R_0.
    
    \end{cases}
\end{equation}
Requiring $2\Omega_K = -\Omega_G$, we then define the virial mass
\begin{equation}
    M_{\rm{vir}}(r) = \begin{cases}
    \frac{5 R_0 \sigma_v^2(r)}{G} x  \ \rm{for} \ r < R_0 \\
    
    {{3 R_0 \sigma_v^2(r)}\over G} {{\Pi_1(r)^2}\over{\Pi_2(r)}}  \  \rm{for} \ r \geq R_0,
    
    \end{cases}
\end{equation} 
and finally the virial parameter  as
\begin{equation}
\alpha_{\rm{vir}}(r) = \begin{cases}
\frac{15 \sigma_v^2(r)}{4 \pi \rho_c G R_0^2} \frac{1}{x^2} \ \rm{for} \ r < R_0 \\
\frac{3 \sigma_v^2(r)}{\pi  \rho_c G R_0^2} \frac{\Pi_1(r)}{\Pi_2(r)} \  \rm{for} \ r \geq R_0.
\end{cases}
\end{equation}

We cast this profile in terms of $f_{DG}$ as defined in Equation \ref{eqn:fdg_Gen}, such that
\begin{equation}
    f_{DG} = 
    \begin{cases}
    {\color{black}1 - \left(\frac{3-k}{3}
    \frac{(R_{CO}/R_0)^3}{ (R_{H_2}/R_0)^{(3-k)} - \frac{k}{3} } \right) \  \rm{for} \ R_{CO} < R_0} \\[1em]
    
    1 - \left(\frac{(R_{CO}/R_0)^{(3-k)} - \frac{k}{3}}{(R_{H_2}/R_0)^{(3-k)} - \frac{k}{3}}\right) \  \rm{for} \ R_{CO} \geq R_0,
    \end{cases}
\end{equation}
assuming $R_{H_2} > R_0$. {\color{black}$R_{H_2}$ can then be solved for analytically, with $\Sigma(R_{H_2})$, $\alpha_{\rm{vir}}(R_{H_2})$, and related properties following.}

\vspace{1.5em}
\subsection{Plummer Density Profile}\label{ap:plummer}

We consider a clump following a Plummer density profile,
\begin{equation}
    \rho(r) =
    \rho_c\left({1\over{\sqrt{x^2+1}}}\right)^\eta,
\end{equation}
where $\rho_c$ is the central density, $x= r/R_0$, $R_0$ is the radius of the central core, and $\eta$ is the index of the power-law {\color{black}at large radii}.  {\color{black}From Equation \ref{eqn:Mr_gen}}, the mass within $r$ is 
\begin{equation}
        M(r) = \pi\rho_c R_0^3 P_1(r),
\end{equation}
where 

\begin{equation}
P_1(r) =
    \begin{cases}
    {\color{black}4(x - \arctan(x)) \ \rm{for} \ \eta = 2} \\
    {\color{black}2(\arctan(x) - {x\over{x^2+1}}) \ \rm{for} \ \eta = 4.}
    \end{cases}
\end{equation}
{\color{black}From Equation \ref{eqn:omegaG},} the gravitational term can be written as
\begin{equation}
    \Omega_G(r) = -\pi^2G\rho_c^2 R_0^5 P_2(r) = -{{GM(r)^2}\over{R_0}} {{P_2(r)}\over{P_1(r)^2}},
\end{equation}
where 

\begin{equation}
P_2(r) =
    \begin{cases}
    {\color{black}16\left(x - \arctan(x) - \displaystyle\int_0^x \frac{x' \arctan(x')}{x'^2+1} \rm{dx'} \right)\ \rm{for} \ \eta = 2} \\[0.8em]
    
    {\color{black}\arctan{(x)} + \frac{x-4\arctan{(x)}}{x^2 +1} + \frac{2x}{(x^2+1)^2} \ \rm{for}   \ \eta = 4.}
    \end{cases}
\end{equation}
{\color{black}From Equation \ref{eqn:omegaK},} the kinetic term is 
\begin{equation}
    \Omega_K(r) = \frac{3}{2} \pi \rho_c R_0^3\sigma_v^2(r) P_1(r).
\end{equation}
We can then derive the virial mass through requiring $2\Omega_K = -\Omega_G$,
\begin{equation}
 M_{\rm{vir}}(r) = {{3R_0  \sigma_v^2(r)}\over G} {{P_1(r)^2}\over{P_2(r)}},
\end{equation}
and finally the virial parameter
\begin{equation}
    \alpha_{\rm{vir}}(r) = {{3\sigma_v^2(r)}\over{\pi \rho_c G R_0^2 }} {{P_1(r)}\over{P_2(r)}}.
\end{equation}

We adapt this profile into $f_{DG}$  as defined in Equation \ref{eqn:fdg_Gen}, such that
\begin{equation}
    f_{DG} = 
    \begin{cases} 
    {\color{black}1- \left(\frac{\frac{R_{CO}}{R_0} - \arctan(R_{CO}/R_0) }{ \frac{R_{H_2}}{R_0} - \arctan (R_{H_2}/R_0)}\right) 
    \ \rm{for}   \ \eta = 2} \\[1.5em] 
    
    1 - \left(\frac{  \arctan(R_{CO}/R_0) - \frac{(R_{CO}/R_0)}{(R_{CO}/R_0)^2 + 1} }{ \arctan(R_{H_2}/R_0) - \frac{(R_{H_2}/R_0)}{(R_{H_2}/R_0)^2 + 1} }\right) \ \rm{for}   \ \eta = 4. 
    \end{cases}
\end{equation}
$R_{H_2}$ can then be solved for numerically to obtain estimates of $\Sigma(R_{H_2})$, $\alpha_{\rm{vir}}(R_{H_2})$, and related properties.  

\bibliographystyle{yahapj}
\bibliography{refs.bib}

\end{document}